# Competing Effect of Biquadratic and Heisenberg Coupling on Magnetic Tunnel Junction Molecular Spintronics Devices

Andoniaina Mariah Randriambololona, Hayden Brown, Eva Mutunga, Andrew Grizzle, Christopher D'Angelo, Pawan Tyagi
University of the District of Columbia, Department of Mechanical Engineering

## Abstract

Heisenberg Exchange Coupling (HC) and Biquadratic Exchange Coupling (BQC) are present in magnetic tunnel junctions (MTJ) and nanoscale elements-based spintronics[1]. MTJ-based molecular spintronics (MTJMSD) offer an unprecedented new way to create devices and highly correlated physics phenomena. Molecular spin channels on the exposed edge of MTJ were found to produce strong exchange coupling experimentally and theoretically, and the nature of coupling was believed to be HC [2]. BQC – which leads to the perpendicular alignment of the spin vectors of the adjacent FM electrodes – also occurs via the insulator or molecular nanostructures. Little is known about the competing effects between these two types of interlayer exchange couplings on MTJMSDs. A systematic study was performed using Monte Carlo simulations (MCS) based on a 3D Heisenberg model. The BQC strength was varied for a device with no molecular HC as well as strong parallel and antiparallel molecular HC coupling. The physical and magnetic properties of the MTJMSD were examined. It was found that increasing BQC strength in an MTJMSD with strong parallel and antiparallel molecular HC had little effect on overall device magnetization as HC still ultimately dominated the device magnetization. The effect of BQC on the device's magnetic equilibrium state was also examined through its temporal evolution. Results suggest that when only molecular BQC is present within the device, the device is unable to reach magnetic stability; on the other hand, when HC and BQC are present within the device, HC encourages the device to achieve greater stability. The results ultimately indicate BQC plays a lesser role in overall device magnetization dynamics as it cannot overcome the more substantial magnetization effect produced by the HC. The presence of BQC provides a plausible explanation for the experimentally observed difference in magnetic phase orientations other than parallel and antiparallel states around MTJMSD.

## 1. Introduction

The magnetic tunnel junction (MTJ) device comprises a semiconductor device with a ferromagnet-insulator-ferromagnet tri-layer configuration. Paramagnetic molecules are covalently bonded to the ferromagnet trilayer junction[3] edges to allow for communication between the two ferromagnet (FM) electrodes, resulting in the formation of the molecular spintronics device (MSD)[4]. MTJs are presently used in memory and logic devices and continue gaining more attention in the semiconductor industry because of their energy efficiency, nonvolatility, and scalability[5]. The integration of molecules in the device provides the promising potential to advance the next generation of these devices due to their ability to utilize electron charge and spin properties and leverage their spin degrees of freedom[4].

Heisenberg exchange coupling (HC), the alignment of electron spins in a parallel and antiparallel configuration, has long been found to exist in magnetic trilayers[6]. The energy expression for HC is written as

$$E_{int} = -J_1(m_1 \cdot m_2) \tag{1.1}$$

, where $J_1$ is the Heisenberg coupling (HC) strength and $m_1$ and $m_2$ are the unit magnetization vectors of the first and second magnetic layers[6], respectively. The existence of biquadratic exchange coupling (BQC) was first found experimentally in the exchange interaction of the bcc Fe/Cr/Fe multilayer[7] and was defined by the second-order Heisenberg interaction:

$$E_{int} = -J_2(m_1 \cdot m_2)^2 \tag{1.2}$$

, where $J_2$ is biquadratic coupling (BQC) strength[6]. The combined energy expression can be expressed in the following form:

$$E_{int} = -J_1 \cos\theta - J_2 \cos^2\theta \tag{1.3}$$

, where $\theta$ is the angle between the magnetic vectors of the two FM electrodes. A positive $J_1$ corresponds to ferromagnetic coupling while a negative $J_1$ represents antiferromagnetic coupling. The biquadratic coupling term $J_2$ is always positive because a positive value lowers the energy when the angle between the magnetizations is perpendicular[8].

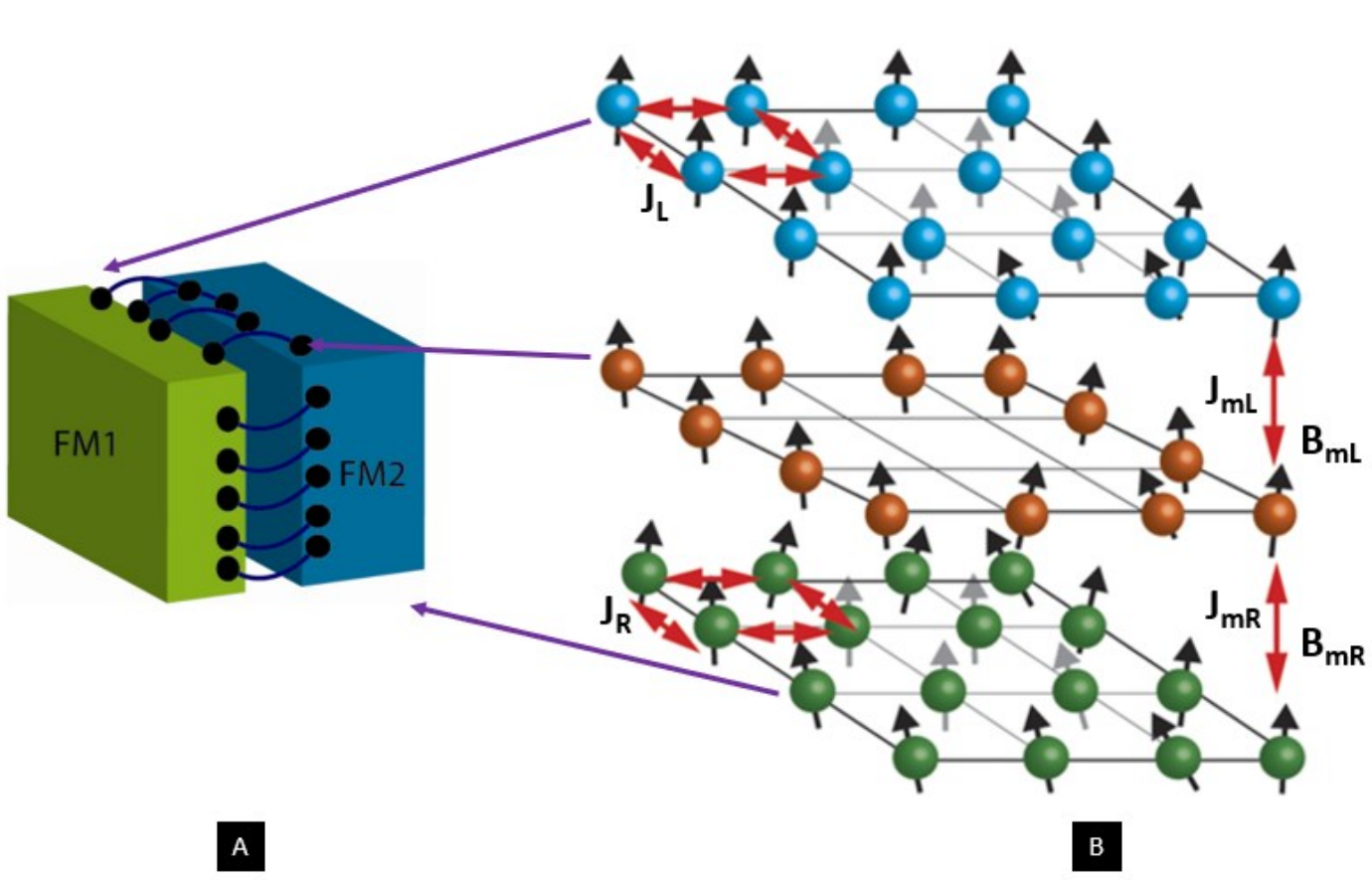

**FIG 1.** (A) MTJMSD with left FM electrode[4] (FM1) and right FM electrode (FM2) with molecules attached to the exposed edges and (B) 3-D molecular representation of the different exchange coupling that exist within the simulations of the device.

Since the discovery of the higher-order exchange coupling [9], extensive research has been done to examine the presence and role of BQC on layered magnetic systems. Bilinear and biquadratic exchange coupling was found experimentally in Fe/Cu/Fe trilayers as a function of interlayer thickness[8] using ferromagnetic-resonance (FMR) and surface magneto-optical Kerr-effect (SMOKE). It was found that the BQC strength increases with terrace width.

Layadi[10] examined the effect of biquadratic coupling and in-plane anisotropy on the resonance modes of an FM-nonFM-FM trilayer. His results concluded that given a positive $J_1$, an additional $J_2$ (where $J_2<0$) increased (decreased) optical (acoustic) mode intensity, while a negative $J_1$ with an increased $J_2$ results in increased (decreased) acoustic (optical) mode while the resonant frequency of both modes decrease[10].

Gareev et al.[11] conducted an experiment that revealed the presence of antiferromagnetic interlayer exchange coupling of Fe layers across epitaxial, Ge-containing spacers and found that exchange coupling changes with spacer thickness from a parallel orientation of electron spins to $90^o$ alignment for Ge wedges and $180^o$(antiferromagnetic) orientation for Si-Ge multilayers, respectively. They deduced that bilinear exchange coupling was comparable for both cases but the addition of BQC was subdued for the Si-Ge multilayer spacers.

Layadi[12] studied the effect of the exchange anisotropy field and HC and BQC coupling strengths on resonance mode behavior for arbitrary HC and BQC parameters and found that in the low magnetic coupling/high exchange anisotropy case, the multilayer system behaved as two uncoupled layers with magnetic characteristics that varied from the initial layer. He found that the effect of low coupling is to modify the different anisotropies while the bilinear exchange coupling contributes to the exchange anisotropy and $J_2$ alters the magnetocrystalline anisotropies. Layadi[13] also compared the analytical expressions for the different switching fields of an MTJ-like system for both weak and strong magnetic interactions to the exchange anisotropy effect and found that the bilinear coupling contributes to the exchange anisotropy fields while the biquadratic coupling effect is equivalent to an in-plane magnetic anisotropy.

Kartsev et al.[1] determined the role of BQC in the magnetic properties of two-dimensional van de Waals layered magnets using the analytical techniques non-collinear first-principles methods and Monte Carlo Simulations (MCS). They discovered that BQC plays a key role in increasing the thermal stability of the layers by intrinsically renormalizing certain magnetic properties such as magnetic anisotropies and spin-wave gaps.

Although extensive research has been done on the competing effect of BQC on MTJs, little is still known about this phenomenon on MSD-based MTJs. Traditional MTJs employ an insulator or non-metal between two FM electrodes, but the addition of paramagnetic molecular channels along the exposed edges has been shown to produce strong coupling between the FMs when competing exchange coupling with the insulator[14]. Prior studies on MTJMSDs have also observed that paramagnetic molecules covalently bonded to two FM electrodes primarily induced strong antiferromagnetic coupling[2, 14, 15].

To better understand the impact of HC and BQC on the MTJMSD, a systematic study using Monte Carlo Simulations (MCS) based on a 3D Heisenberg model was employed. The first set of simulations performed involved varying the molecular BQC for a device with (i) no molecular HC (ii) strong parallel molecular HC and (iii) strong antiparallel HC. The second set of simulations explored the effect of temperature on an MTJMSD with varied molecular HC and BQC strengths.

## 2. Computational Methods

### 2.1 Monte Carlo Simulation Model

In this study, Monte Carlo simulation (MCS)[16] is employed to examine device magnetization behavior. The 3D Ising model with nearest-neighbor interactions is also integrated with the simulation to analyze the spin states of the MTJMSD[17]. The Ising model[18], which determines the total energy of the system, is defined by the standard Hamiltonian equation:

$$\mathcal{H} = -J \sum_{neighbors} S_i \cdot S_{i+1} \quad (2.1)$$

, where J is the exchange coupling between the interacting neighbors, and $S_i$ and $S_{i+1}$ are the atomic spins of the two interacting neighbors of interest. The continuous model is also integrated with the MC simulation to allow the spin vectors to settle in any randomly selected direction according to the equilibrium energy established by the Hamiltonian equation[3]. The standard Hamiltonian in equation 2.1 can be expanded to the following form to include all exchange couplings that were examined in this study:

$$\mathcal{H} = -J_L \left(\sum s_i \cdot s_{i+1}\right) - J_R \left(\sum s_i \cdot s_{i+1}\right) - J_{mL} \left(\sum s_i \cdot s_{i+1}\right) - J_{mR} \left(\sum s_i \cdot s_{i+1}\right) - b_{mL} \left(\sum s_i \cdot s_{i+1}\right) - b_{mR} \left(\sum s_i \cdot s_{i+1}\right) \quad (2.2)$$

, where $J_L$ and $J_R$ are the HC strengths between the left and right FM electrode molecules respectively, $J_{mL}$ and $J_{mR}$ are the HC strengths between the molecules and the left and right FM electrodes respectively, and $b_{mL}$ and $b_{mR}$ are the BQC strengths between the molecules and the left and right FM electrodes respectively. A visual representation of the various exchange interactions that exist in this study is displayed in Fig. 1.

The HC ($J_{mL}$ and $J_{mR}$) and BQC ($b_{mL}$ and $b_{mR}$) strength between the molecules and FM electrodes are varied for the two studies to investigate their competing effects on the MTJMSD magnetization behavior. After providing specific values of the HC and BQC in the system, the Markov chain scheme, and the metropolis algorithm randomly walks through the phase space to minimize device energy and generate a new state. The Marvok process produces a new state by using the transition probability $\tau(s \to s')$ such that it occurs with a probability of the equilibrium Boltzmann distribution:

$$P_{eq}(s) = Z^{-1} \exp\left[\frac{-\mathcal{H}(s)}{kT}\right] \quad (2.3)$$

, where Z is the partition function, s is the spin state, and k and T are the Boltzmann constant and temperature respectively. The electron state is expected to occur with probability $P_k(s)$ at the *k*th time step, as described in the master equation:

$$P_{k+1}(s) = P_k(s) + \sum_{s'} [\tau(s' \to s) P_k(s') - \tau(s \to s') P_k(s)] \quad (2.4)$$

, where $\tau$ is the transition probabilities. Note that the sum is over all possible states, s'. The first term in the sum describes all processes reaching state s, while the second term describes all processes leaving states.

The Metropolis algorithm is what then determines whether to accept or reject the new spin state derived from the energy equation:

$$E = -J_L(\Sigma_{i\in L}\vec{s}_i\vec{s}_{i+1}) - J_R(\Sigma_{i\in R}\vec{s}_i\vec{s}_{i+1}) - J_{mL}(\Sigma_{i\in L,i+1\in mol}\vec{s}_i\vec{s}_{i+1}) - J_{mR}(\Sigma_{i-1\in mol,i\in R}\vec{s}_{i-1}\vec{s}_i) - b_{mL}(\Sigma_{i\in L,i+1\in mol}\vec{s}_i\vec{s}_{i+1})^2 - b_{mR}(\Sigma_{i-1\in mol,i\in R}\vec{s}_{i-1}\vec{s}_i)^2 \quad (2.5)$$

Under this algorithm, the spin vector direction of a randomly selected site is changed to produce a new spin vector state[16] with a magnitude of 1. The energy equation is used to calculate the new and old energy states. If the difference between two energy states is less than zero or $\exp[-\Delta E/kT] \geq r$, where r is a uniformly distributed random variable with a range from 0 to 1, then the new spin state will be accepted[2, 16].

## 2.2 Simulation Condition for Study 1

In the first study, the molecular BQC strength is varied between 0 and 1 for a device with three different fixed HC conditions: (i) no molecular HC (JmL/JmR = 0/0) (ii) strong par2allel molecular HC (JmL/JmR = 1/1) and (iii) strong antiparallel HC (JmL/JmR = -1/1). To understand whether the additional FM mass from the extended leads of the MTJMSD influences the total magnetic behavior of the device, a pillar configuration of the MTJMSD is also examined. The extended cross junction and pillar configurations are illustrated in Fig. 2(A) and 2(B), respectively.

In the simulations, for the MTJMSD with extended leads, the two FM electrodes, labeled left and right, are oriented perpendicular to one another to form a cross junction and are each of 5x5x50 dimensions. These dimensions were chosen to ensure that the MCS results are repeatable while also maintaining a relatively realistic resemblance to results from experimental studies[16]. The total number of molecules in the left and right electrodes of the MTJMSD with extended cross junctions equates to 2,500 (which means there exists 1,250 molecules per FM electrode).

The paramagnetic molecules are bonded around the edges of the tunnel junction to connect the left and right FM electrodes and are represented by a square-shaped perimeter with a 5x5 area. In this study, only the two side molecular edges are turned on. The tunnel barrier is also represented by an empty interior inside the molecule parameter as displayed in Fig. 2(A). The two FM electrodes are coupled via the molecules while there is no communication between the two bodies via the empty space. This is because the empty space is defined to act like a perfect insulator in the study (In the simulation, this means the exchange interaction between the left and right electrode molecules, JLR = 0).

For the pillar configuration, the same simulation conditions specified for the MTJMSD with the extended cross junctions is applied. The only difference is the dimension of the left and right electrodes, which are of 5x5x5 dimensions, for a combined 250 total molecules. The total number of molecules in the electrodes is of importance for the analysis of the temporal evolution graphs, which will be discussed later in the paper.

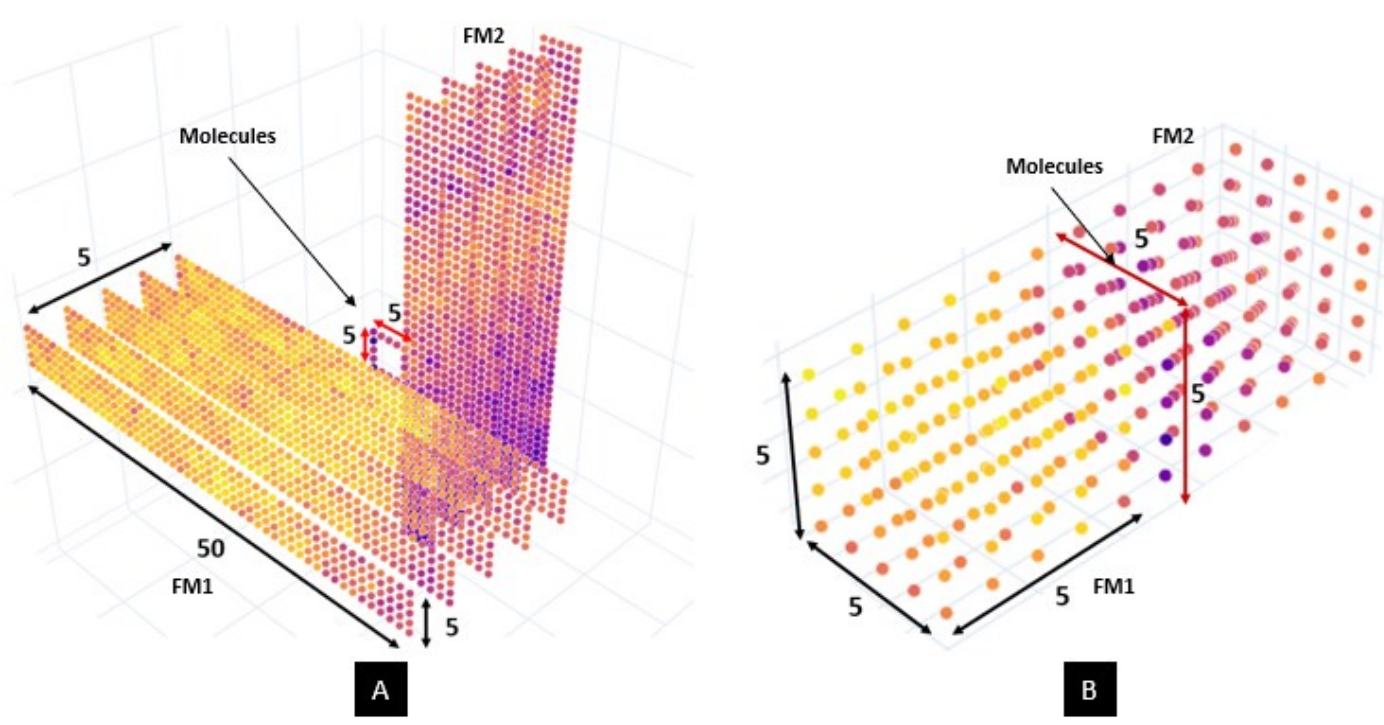

**FIG 2.** Lattice plots of (A) MTJMSD with extended cross junction and (B) MTJMSD pillar structure. Both configurations are comprised of a FM-molecules-FM structure with 5x5 molecules around the edges. In these specific simulations, both the top and bottom edges are turned off.

For all MC simulations, the boundary conditions were selected in a way where the atomic spins beyond the boundary atom of the MTJMSD are zero[2]. Thermal energy (kT) is kept at a constant 0.1, which encompasses a practical working temperature range for devices since kT=0.1 is equivalent to the 50-130 degree range[3]. The molecular BQC coupling strength between the molecules and the left and right FM electrodes and the molecular HC strength between the two FM electrodes are varied. No external magnetic field is applied in the simulations.

Three different types of HC interactions between the molecules and the left and right electrodes were examined: (i) one with no HC between the molecules and the left and right electrodes (JmL/JmR = 0/0) (ii) parallel molecular HC (JmL/JmR = 1/1) and (iii) antiparallel HC ($J_{mL}/J_{mR}$ = -1/1). For each HC condition, the BQC between the molecules and the left and right electrodes was varied between 0 and 1 in increments of 0.1. To achieve a stable low energy state, 200-500 million iterations are run in the MCS. 3 iterations of each of the BQC conditions were performed to ensure repeatability of results.

## 2.3 Simulation Conditions for Study 2

For the second study, only the pillar MTJMSD was simulated for. The HC for three different thermal energy (kT) conditions (kT = 0.05, 0.1, and 0.2) is varied for a large range of coupling strengths between the magnitudes of -1 and 1 and a weak (bmL/bmR = 0.05/0.05) and strong (bmL/bmR = 0.5/0.5) molecular BQC strength is applied to the device.The pillars in this set of simulations have the same dimensions as the pillars in section 2.2, except all four molecular edges are turned on. The simulation conditions are also the same as section 2.2 except a much wider range of HC pairs were examined.

## 3 Results

### 3.1 Condition 1: Varying BQC strength with no HC

Under condition 1, molecular BQC strength is varied while no HC is present in the molecules (JmL/JmR = 0/0). Temporal evolution graphs were used to visualize the total magnetic behavior over time. Fig. 3 depicts the temporal evolution graphs for both the extended cross junction MTJMSD and the pillar MTJMSD for no, moderate, and strong BQC interactions. The temporal evolution graphs provide insight on whether the device achieves an equilibrium magnetic state given the specified number of iterations. ML norm and MR norm are the magnetic moments of the left and right FM electrodes, respectively. They stabilize when their respective lines reach steady state at around 1,250 for the extended cross junction configuration and around 125 for the pillar configuration. If the overall equilibrium magnetic moment (M norm) of the two electrodes reaches close to the total number of FM electrode molecules (2,500 total for the MTJMSD with the extended cross-junction and 250 total for the pillar MTJMSD), the device behaves ferromagnetically; The total device stabilizes antiferromagnetically if the M norm reaches close to zero.

As shown in Figs. 3(A) and 3(D), when no molecular HC and BQC exists in the device, no relation exists between the left and right electrode as the overall equilibrium magnetic moment never stabilizes to a specific magnetic behavior for both the extended cross junction case (Fig. 3(A)) and the pillar case (Fig. 3(D)); instead the two electrodes behave as separate bodies that have their own spin behavior, which is attributed to their separate FM property ( where JL = 1 and JR = 1 in the simulations). In this instance, the molecules behave as an insulator and the two FM's spins are decoupled with no communication between their spins.

Introducing BQC, especially at a moderate strength of BQC does assist the device in establishing some form of device magnetic stabilization. Figs. 3(B) and 3(E) depict the general magnetic behavior of the extended cross junction and pillar configuration, respectively for BQC strength 0.5/0.5. For the extended cross junction case, the overall equilibrium magnetic state of the device is close to completely ferromagnetic; on the other hand, the pillar MTJMSD is leaning more towards and antiferromagnetic state. This inconsistent and random orientation of magnetic state has been the case when increasing BQC above 0.4/0.4 and when comparing the magnetic behavior of the two MTJMSDs. For instance, when examining the evolution of the overall magnetic behavior of the MTJMSD as BQC is increased from 0.5 (Fig. 3(B)) to 1 (Fig 3(C)), the total device magnetization flipped from ferromagnetic to antiferromagnetic; the same switching behavior was exhibited for the pillar (Refer to Figs. 3(E) and 3(F)). The results indicate that there is a high probability of the spin either orienting mostly ferromagnetically or antiferromagnetically especially at high (bmL/bmR < 0.5/0.5) BQC strengths, because BQC has no preference for parallelity. If the spins of the molecules are oriented perpendicular with respect to the two electrodes' spins, the device will be biquadratically coupled.

The results from the device with the extended cross junction also posits that the additional spins from the extended leads does have some role in device magnetic stabilization because device tends to stabilize to a specific magnetic behavior much quicker than the pillar device even at small BQC strengths.

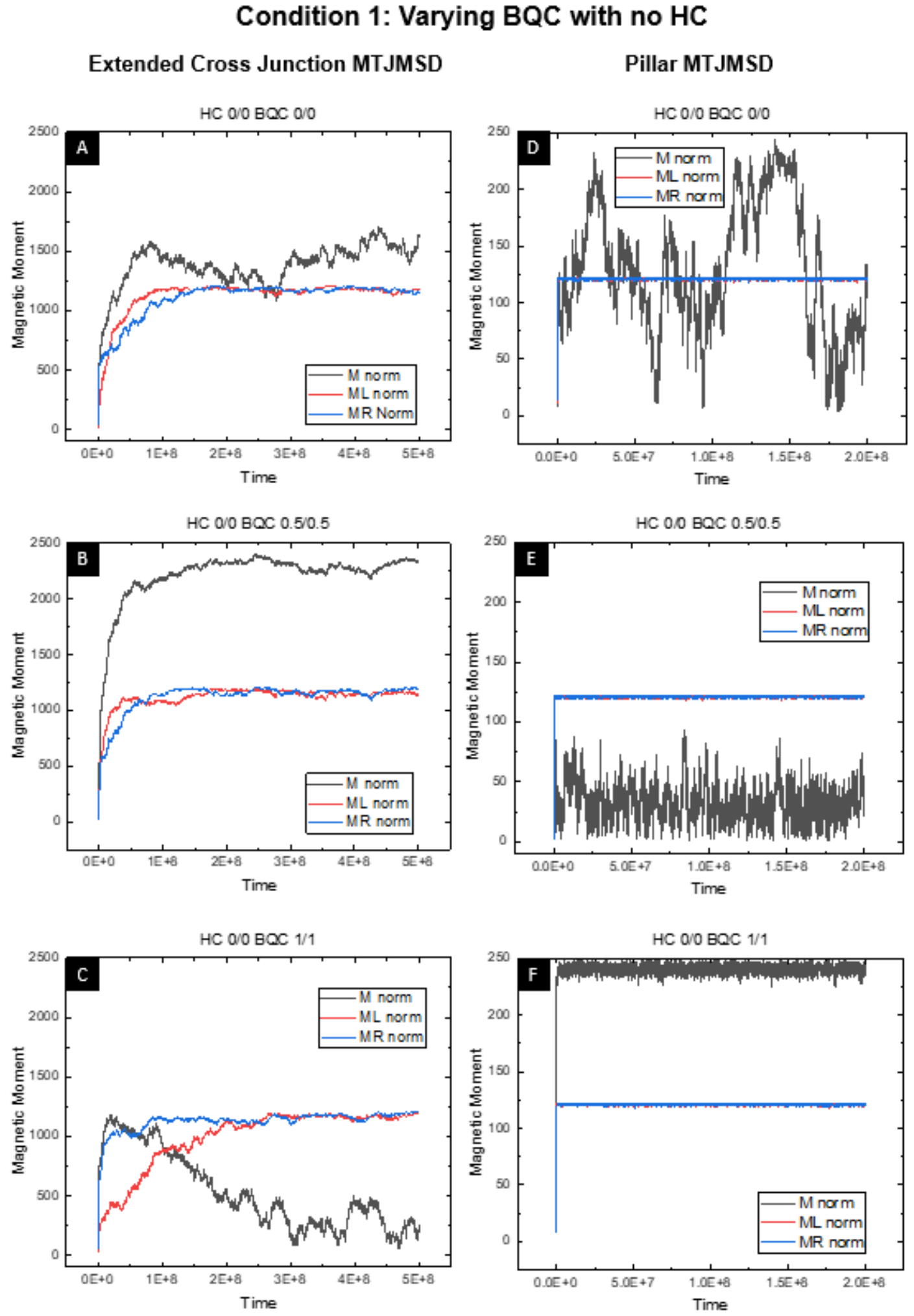

**FIG 3.** Condition 1 temporal evolution graphs for (a)no (BmL/BmR 0/0) (b) moderate (BmL/BmR 0.5/0.5) and (c) strong (BmL/BmR 1/1) BQC in an MTJMSD with extended cross junctions and (a)no, (b)moderate and (c) strong BQC coupling in an MTJMSD with pillar MTJMSD, respectively. For all simulations under this condition, no HC exists in the molecules.

The extended cross junction and pillar MTJMSD also exhibit similar total device magnetic behaviors when they are in a ferromagnetic or antiferromagnetic state. For both the extended cross junction and pillar cases, the devices were able to easily reach ferromagnetic state (Figs. 3(B) and 3(F)), but struggled to achieve antiferromagnetic coupling (Figs. 3(C) and 3(E)).This behavior agrees with the experimental fact that antiparallel magnetic resistance will always be larger than parallel resistance [19].

The autocorrelation graphs for this condition were also examined to understand how correlated the spins of the molecular channels are to the left and right FM electrodes. The autocorrelation graphs only correspond to the MTJMSD with extended leads. As illustrated in Fig. 4(A), under no HC and no BQC, the molecule spins are correlated with one another, but the spins of the electrodes do not align with the spins of the molecules; instead, they randomly orient in their own magnetic domains from a correlation range of -0.5 to 0.7 with respect to the molecules. This reinforces the idea that the electrodes have no communication and function as their own separate magnetic bodies.

Figure 4(B) depicts the evolution of the magnetic behavior at BQC strength 0.5/0.5. The alignment of the spins in the electrodes appear to be parallel with one another and the spins in the of molecules appear to be around 45 degrees with respect to the electrodes.

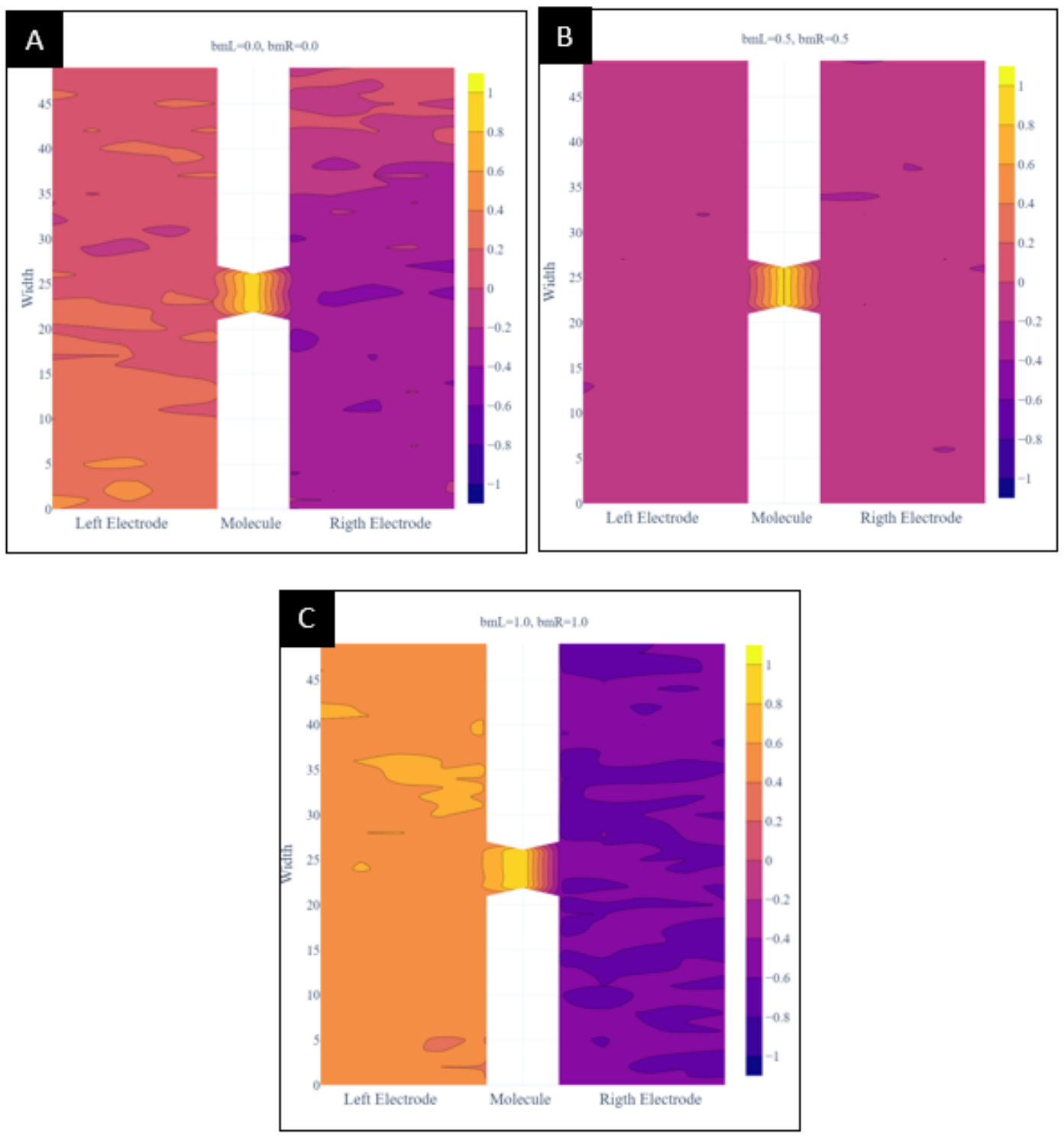

**FIG 4.** 2-D Autocorrelation graphs for (A) no (B) moderate and (C) strong BQC given no molecular HC is introduced to the system. The autocorrelation charts visually depict how correlated the molecules are with respect to the left and right electrodes.

At the highest BQC strength of 1 (Fig. 4(C)), the left and right electrodes assume a more antiparallel state while they are close to a 90-degree orientation with the molecules. The results from the autocorrelation graphs affirm the conclusion that when no HC is present, the device does not favor an alignment of the electrodes if they are mostly perpendicular with respect to the molecules. However, the degree to which the electrodes are perpendicular with respect to the molecules is not entirely dependent of BQC strength since results show that extremely high and low correlations with the molecules do not necessarily occur at the highest BQC strength.

Since condition 1 only examines the effect of BQC on the device, it can be deduced that BQC does not influence the electrode spins to orient in a preferred orientation as it has no preference for parallelity; the strength at which the electrodes align in a parallel or antiparallel is also on the weaker side, which indicates that BQC weakly influences electrode spin in the electrodes.

### 3.2 Condition 2: Varying BQC strength with strong parallel coupling

For condition 2, the device was subjected to a fixed strong parallel HC (JmL/JmR = 1) and BQC strength was varied between 0 and 1. When only HC exists in the device (Figs. 5(A) and 5(D)), the MTJMSD's electrodes are able to communicate via the electrodes, subduing them into a parallel spin state.

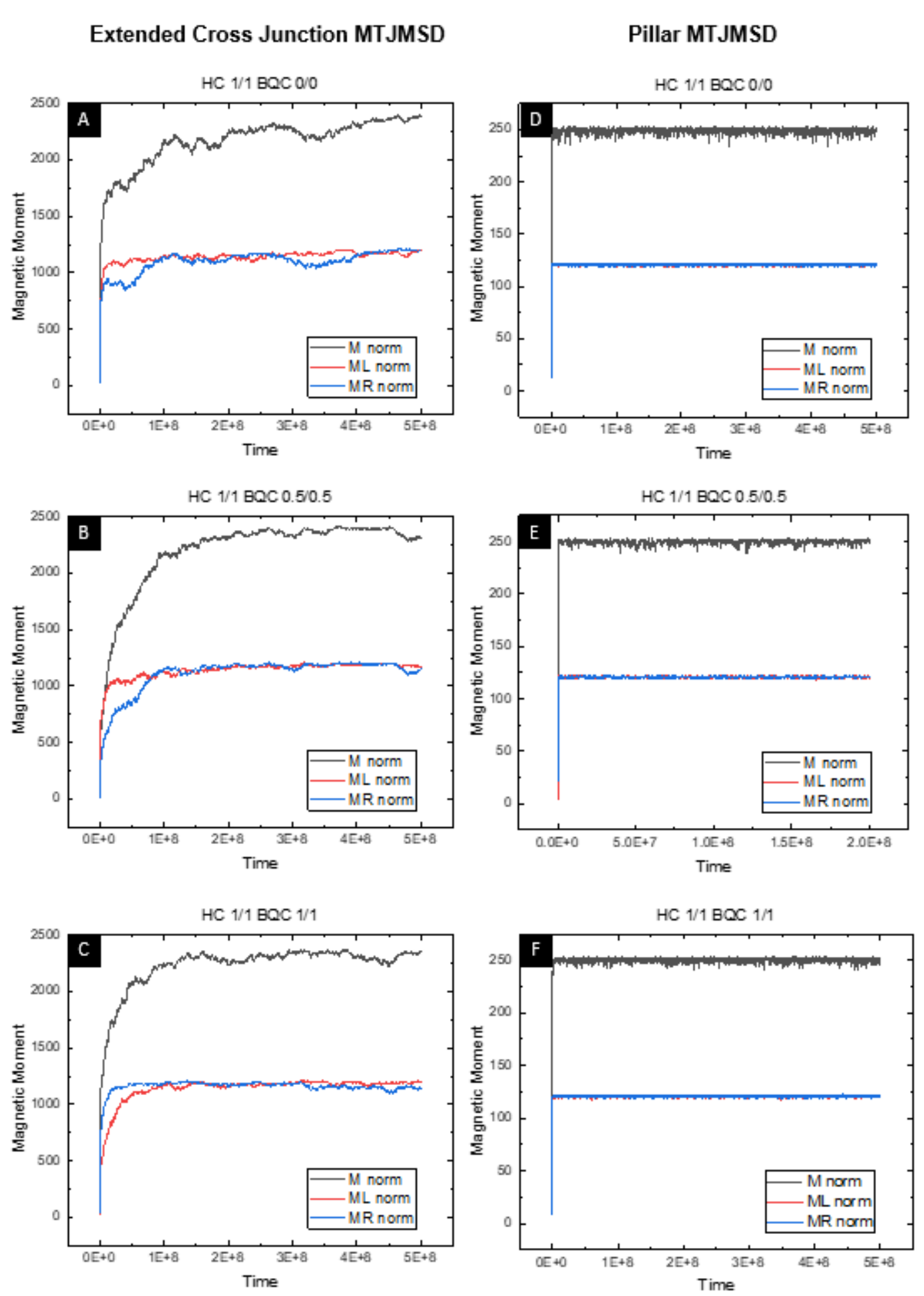

**FIG 5.** Temporal evolutions graphs for condition 2 for the extended cross junction MTJMSD under (A) no, (B) moderate, and (C) strong BQC and the pillar under (D), (E) moderate and (F) strong BQC. Both extended cross junction and pillar MTJMSD simulations under condition 2 exhibit FM behavior.

Introducing and increasing the BQC strength up to the same magnitude as HC was found to have little influence on device magnetization; it only assisted the device in achieving magnetic equilibrium more quickly for the device with extended cross junctions as shown in the comparison between the evolution between BQC = 0/0, BQC = 0.5/0.5 and BQC = 1/1 shown in Figs. 5(A), 5(B) and 5(C), respectively. The spins of the molecules only try to establish a $90^0$ orientation with respect to the left and right electrodes, especially at the highest BQC strength (bmL/bmR = 1/1).

Both the MTJMSD cross junction and pillar simulations exhibit the same FM behavior regardless of increased BQC strength, as shown in the temporal evolution graphs in Fig. 5; however, the pillar MTJMSD reach stability much quicker than the extended cross junctions. This can be attributed to the fact that the extended cross junction has one order of magnitude more molecules to align in the electrodes as compared to the pillar.

The autocorrelation graphs for this condition is displayed in Fig. 6. Curiously, there exists high correlation between the molecule spins and the spins of the left and right electrodes near the molecules for all BQC strengths examined (bmL/bmR = 0/0 to bmL/bmR = 1/1), but some portions of the extended edges have lower correlation with respect to the rest of the device. This resistance against uniformly orienting the spins at the edges with the spins of the rest of the electrode body could potentially be explained by the idea that the effect of BQC is a predominantly localized behavior and is competing with the magnetic moments of the additional magnetic mass from the extended leads.

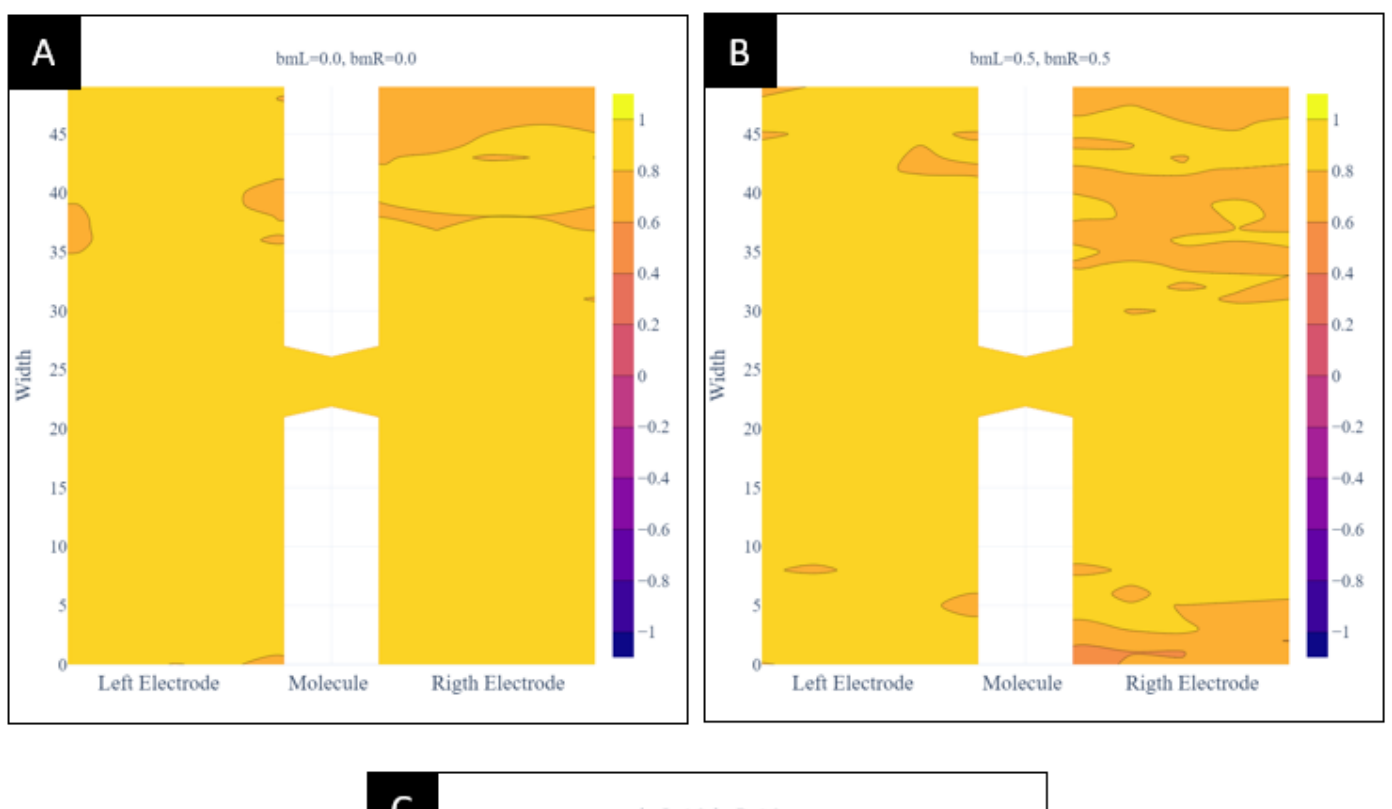

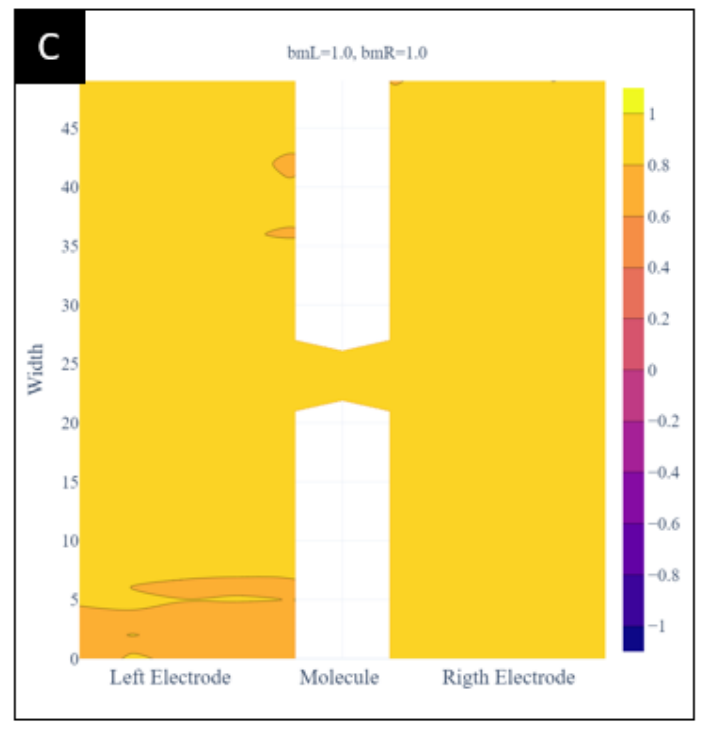

**FIG 6.** 2-D Autocorrelation graphs for (A) no (B) moderate and (C) strong BQC given strong parallel molecular HC

Condition 2 results ultimately conclude that under the condition of high parallel HC, the total device stabilizes quickly, especially under the additional influence of BQC. This behavior can be explained by the fact that maintaining a parallel spin state involves little resistance[19], but this behavior can also be attributed to the addition of increased BQC strength.

### 3.3 Condition 3: Varying BQC strength with strong antiparallel coupling

The final case under the first set of simulations has the devices subjected to a fixed strong antiparallel HC while BQC was varied between 0 and 1. Much like condition 2, the (antiparallel) HC dominated the overall magnetization behavior regardless of increased BQC strength, as illustrated in the temporal evolution graphs in Fig. 7 below.

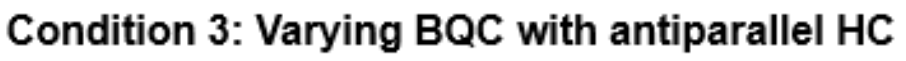

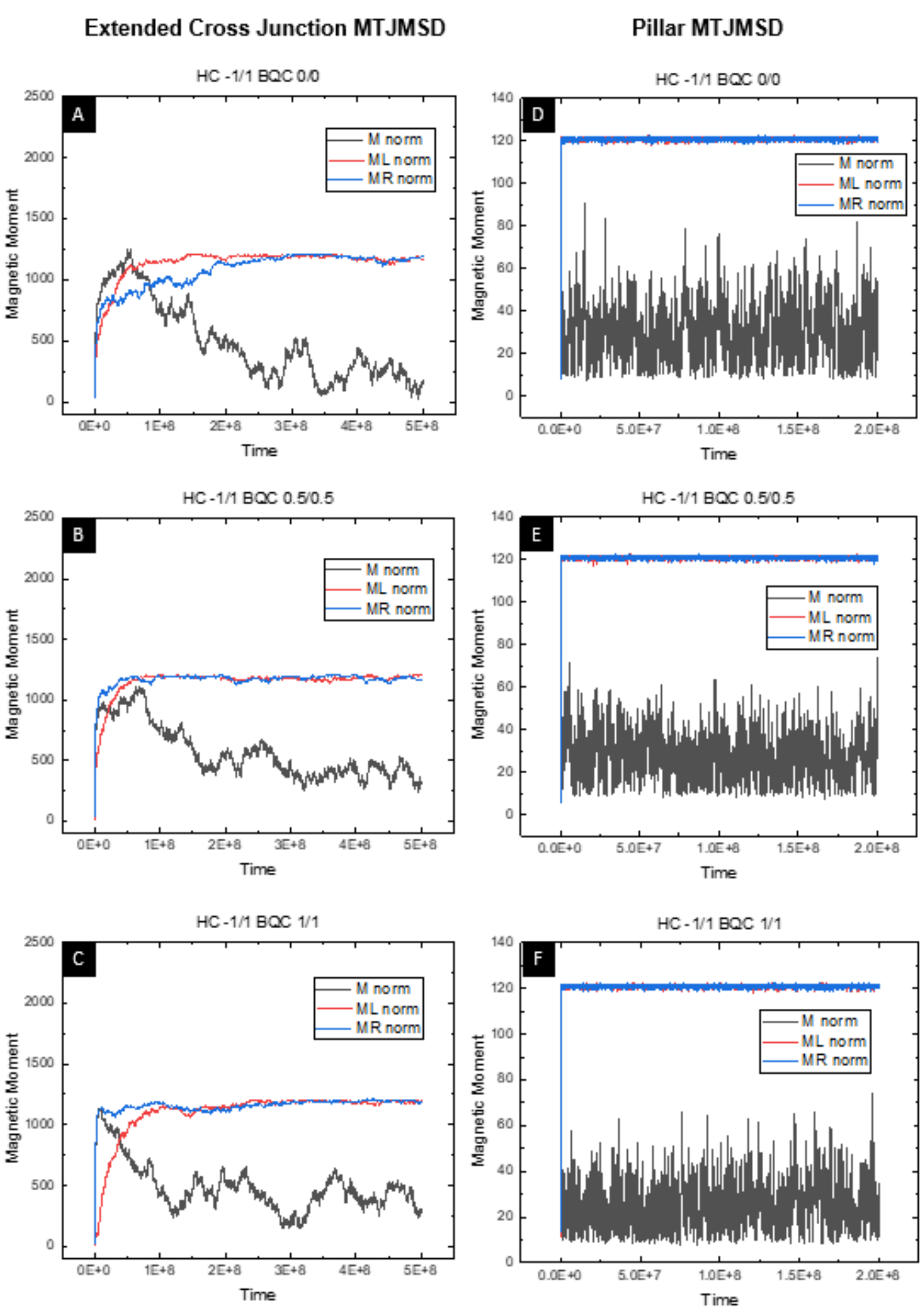

**FIG 7.** Temporal evolutions graphs for condition 3 for the extended cross junction MTJMSD under (A) weak, (B) moderate, and (C) strong BQC and the pillar under (D) weak, (E) moderate and (F) strong BQC. Both extended cross junction and pillar MTJMSD simulations under condition 3 exhibit FM behavior.

The overall equilibrium magnetic moment behavior given by the temporal evolution graphs of the extended cross junction and pillar MTJMSD agree with one another; All graphs reach an overall equilibrium magnetic moment in the antiparallel state. The difference between the two configurations' magnetic behavior lies in the degree of stability to which the molecules of the left and right electrodes are oriented antiparallel with respect to one another. There is a greater resistance against establishing an antiparallel alignment of the spins in the pillar device compared to the extended cross junction, which posits that the extended leads aid the magnetic stabilization of the device when under the influence of high molecular HC.

Like condition 2, the edges of the leads on the device also experience a mixing in the orientation of spins in the sense that it does not fully correlate with the spin orientation of the entire electrode; instead, it has a weaker correlation with the molecules, which emphasizes that BQC may be a more localized phenomena experienced at the junction to aid in the overall equilibrium magnetic moment stability but not so much as it extends further away from the junction. This conclusion is depicted by the edges in the autocorrelation charts in Fig. 8.

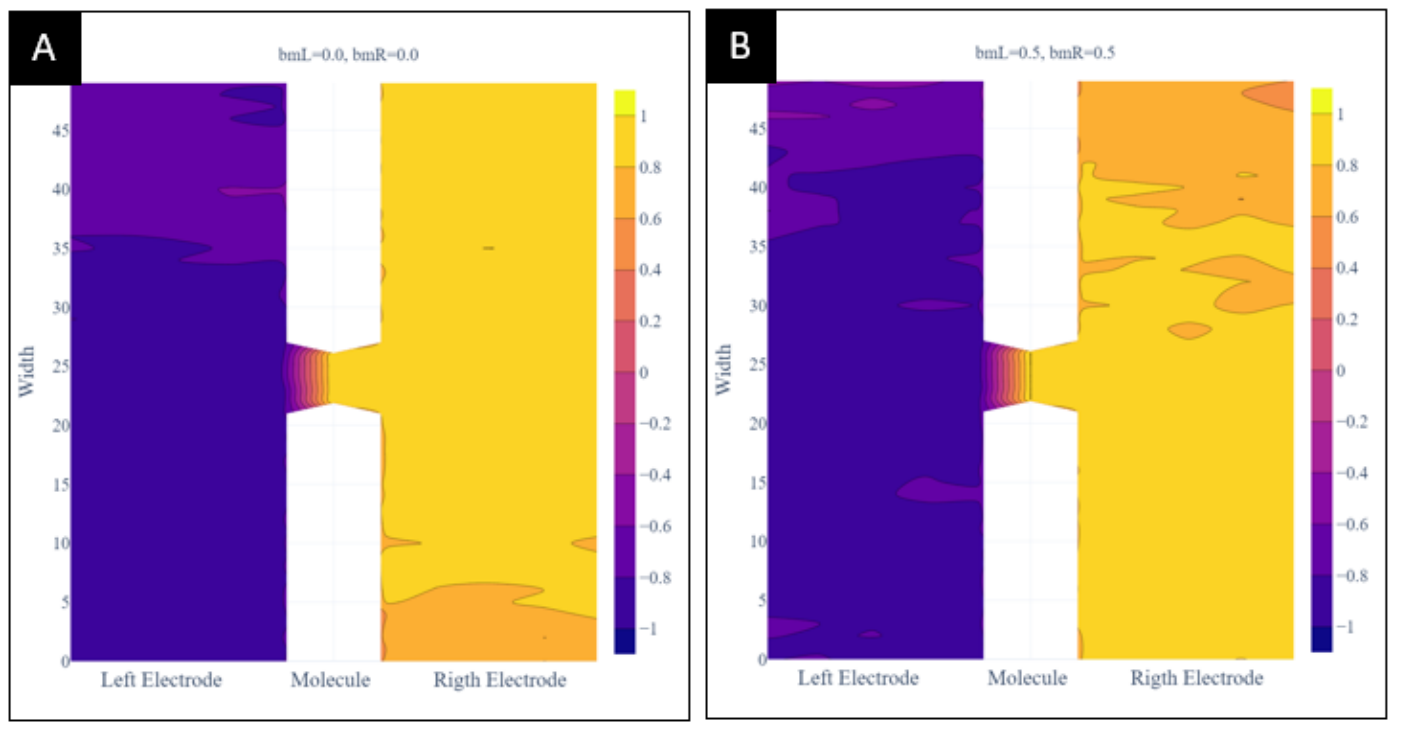

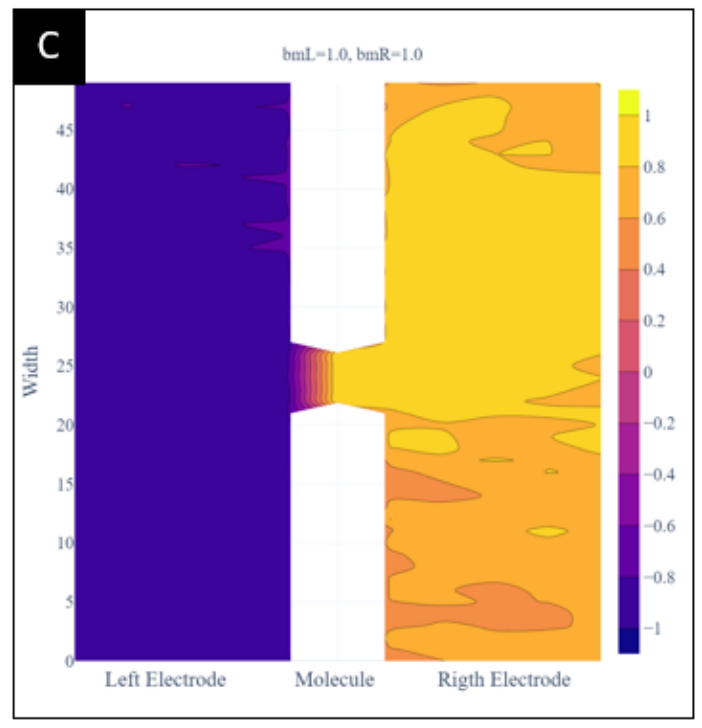

**FIG 8.** Autocorrelation graphs for (A) no (B) moderate and (C) strong BQC given strong parallel molecular HC

The simulation results for conditions 2 and 3 ultimately are consistent with the expected behavior that when parallel (antiparallel) HC exists, especially strong HC (where HC > BQC), the spins tend to orient ferromagnetically (antiferromagnetically). Increased BQC strength for both cases, especially at cases of equal BQC strength to HC strength pushes the molecules to align perpendicular to the left and right FM electrodes. The molecules also do not necessarily compete with HC but rather aids in the stabilization of the device magnetization towards either a parallel or antiparallel configuration.

### 3.4 Thermal effects on device magnetization

Temperature has an apparent effect on the alignment of spins, particularly for the low antiparallel molecular HC cases (in JmL/JmR -/+ and JmL/JmR +/-). The contour plots in Fig. 9 give more insight on this behavior. When considering the effect of temperature given a low BQC case where bmL/bmR = 0.05, increasing kT from 0.05 (Fig. 9(A)) to 0.1 (Fig. 9 (B)) influences the spins in the antiparallel HC cases to align more antiferromagnetically even at low JmL/JmR values. This effect is even greater at kT = 0.2, as shown in Fig. 9 (C), where almost all spins in the JmL/JmR -/+ and JmL/JmR +/- quadrants start to align antiferromagnetically. For the contour plots in Figs. 9(A), 9(B) and 9(C), the quadrants with the parallel HC configurations (JmL/JmR +/+ and JmL/JmR -/-) also exhibit randomizations in their spin states as kT is increased, although there is not as an extreme change in their spin states unlike the low antiparallel HC cases.

This behavior can be explained by the fact that additional thermal energy will excite the electrons to randomly orient themselves until they reach a spin state that is energetically favorable to maintain (i.e., the antiferromagnetic state). This also means that the increased thermal energy reduces the spins' resistance to altering their spin state and will be able to switch state easily. This has important implications when it comes to understanding what temperature the devices should be subjected to ensure that they exhibit desired magnetic behavior for various purposes.

The effect of temperature and BQC coupling is more apparent when comparing the bmL/bmR = 0.05 case with the bmL/bmR = 0.5 case. When a stronger BQC is introduced, BQC attempts to magnetically stabilize the device by reducing the randomness of the spins and even attempting to orient them in a ferromagnetic fashion. This additional behavior of wanting to orient in a particular direction may be a result of all 4 molecular edges being turned on for this set of simulations performed. This phenomenon is depicted by a comparison between Figs. 9(A) and 9(D), where thermal energy kT = 0.05 remains constant for both scenarios but BQC is varied between weak BQC strength (BQC = 0.05/0.05) and strong coupling strength (BQC = 0.5/0.5).

As shown in Figure 9(D), at kT = 0.05, only strong -/+ or +/- JmL/JmR values (where JmL/JmR is greater than bmL/bmR = 0.5/0.5) cause the spins of the left and right FM electrodes to orient antiparallelly with respect to one another. Like the BQC = 0.05 case, increasing the thermal energy kT to 0.1 increases the likelihood for randomness because of the drop in resistance from the device, as shown in Fig. 9(E). Further increasing kT to 0.2 causes increased randomness of the spins until the left and right electrodes align close to or reach 180-degree alignment with respect to each other.

A comparison of weak BQC (bmL/bmR = 0.05/0.05) and strong BQC (bmL/bmR = 0.5/0.05) at kT = 0.05 suggests that the presence of increased BQC strength encourages the antiparallel spins in the JmL/JmR -/+ and JmL/JmR +/- conditions to become parallel when the strength of HC is less than the strength of BQC. The same conclusion can be said for the comparison between Fig. 9(B) and Fig. 9(E), and Fig. 9(C) and Fig.9(F).

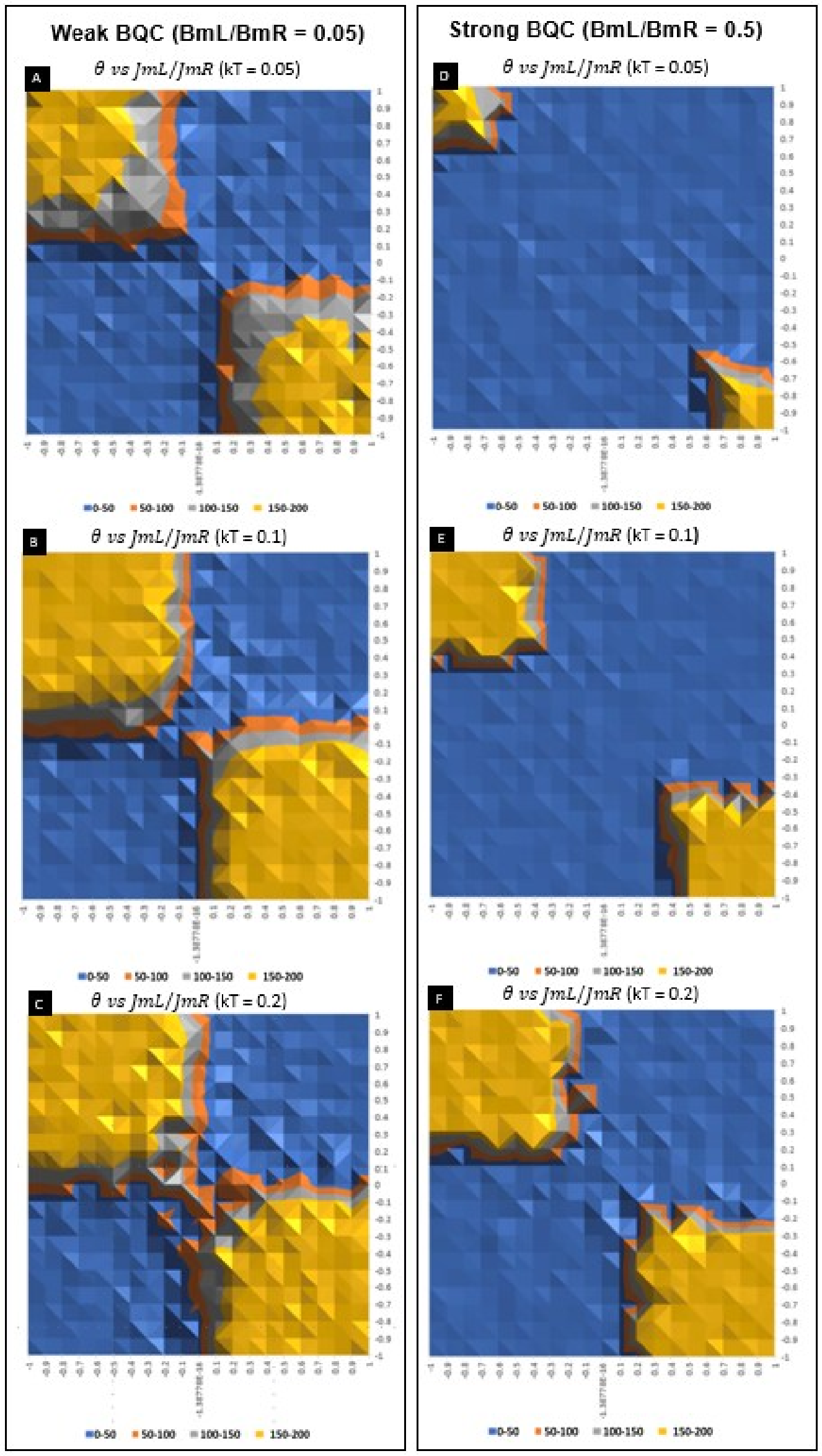

**FIG 9.** Contour plots for weak (bmL/bmR = 0.05/0.5) BQC case for (A) kT = 0.05, (B) kT = 0.1 and (C) kT = 0.2. Contour plots for weak (bmL/bmR = 0.05/0.5) BQC case for (D) kT = 0.05, (E) kT = 0.1 and (F) kT = 0.2. Each contour plot is a function of spin angle $\theta$ between the left and right ferromagnet electrodes and JmL/JmR values. The contour plots are broken into 4 quadrants: JmL/JmR +/+, JmL/JmR -/+, JmL/JmR -/- and JmL/JmR +/-.

The average magnetic behavior of each quadrant for each kT condition was also examined given both weak and strong BQC interactions to quantify the change in spin behavior. Three iterations of each condition were performed to ensure that the average results are repeatable. As shown in Fig 10(A), it was found that for the parallel cases (where JmL/JmR = -/- and +/+), at weaker BQC (bmL/BmR = 0.05), increasing the thermal energy kT from 0.05 to 0.2 resulted in a change in average angle of about $11^0$. When BQC is increased to bmL/bmR 0.5, only a change of about $2^0$ was seen when kT was increased from 0.05 to 0.2. This reaffirms the conclusion that increased temperature encourages randomness of the alignment of spins in the parallel HC case, but the addition of BQC reduces the device's susceptibility to spin fluctuations by encouraging the spins to stabilize back down to its ferromagnetic state.

Interestingly, slightly different conclusions are derived from the antiparallel HC case in Fig. 9(B). for the antiparallel cases (where JmL/JmR = -/+ and +/-), at weaker BQC (bmL/BmR = 0.05), increasing the thermal energy kT from 0.05 to 0.2 resulted in a change in average angle of about $72^0$. When BQC is increased to bmL/bmR 0.5, larger change of about $106^0$ was seen when kT was increased from 0.05 to 0.2. This means that for the antiparallel HC case, increasing thermal energy results in significantly larger changes in the average angle between the left and right electrodes; increasing BQC on the other hand did attempt to reduce thermal fluctuations but the device struggled to overcome the effects of temperature after kT = 0.1.

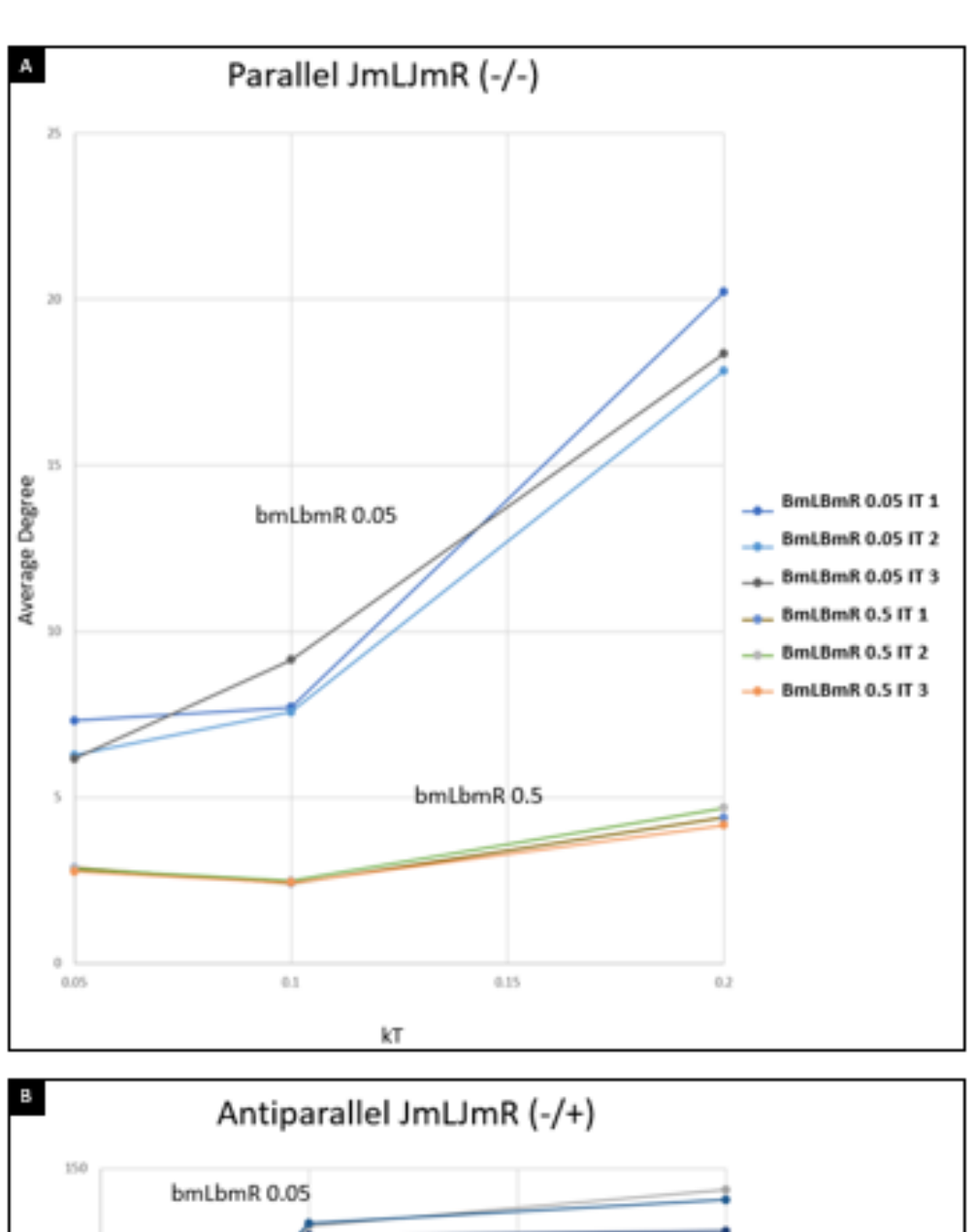

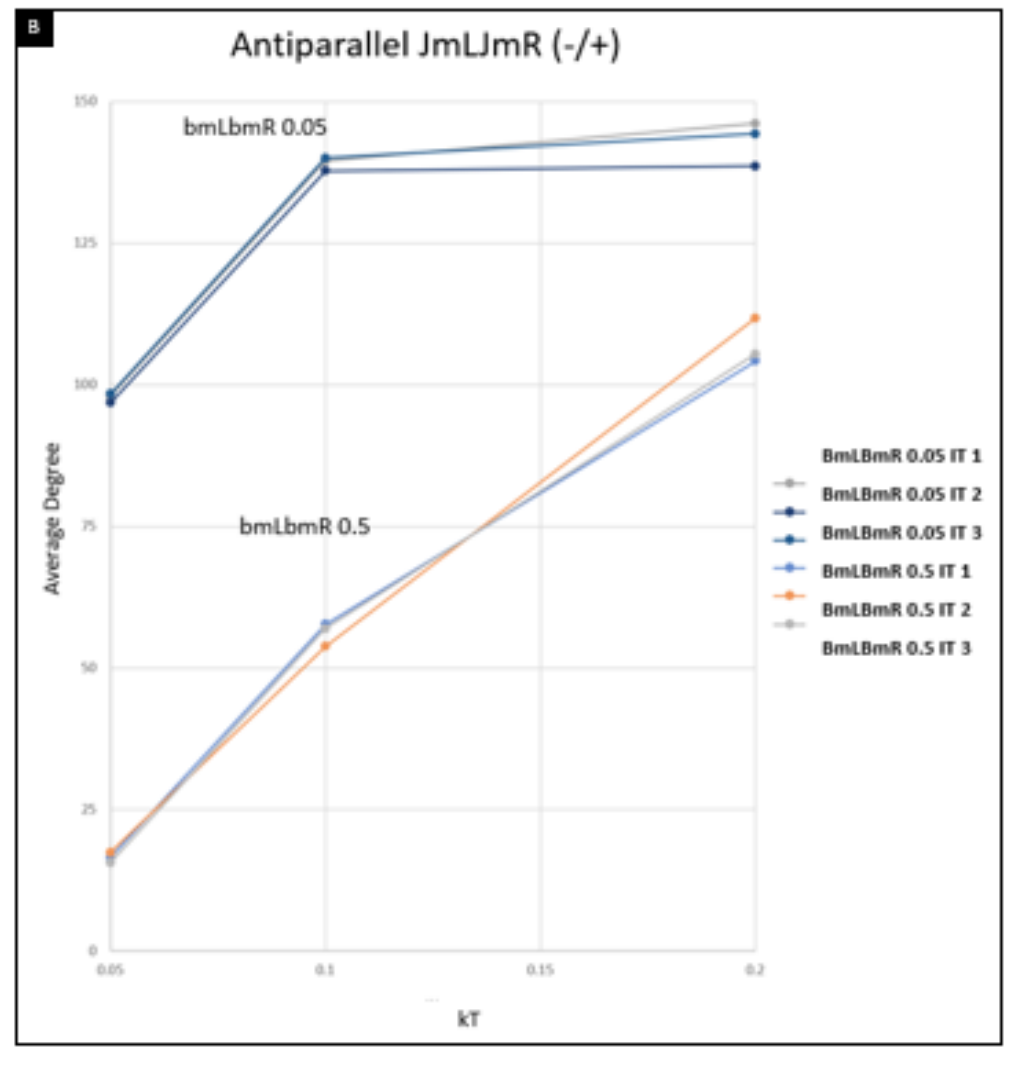

**FIG 10.** Average degree vs kT for (A) parallel HC cases (JmL/JmR -/-) for weak (BQC = 0.05) and strong (BQC = 0.5) interactions and (B) antiparallel (JmL/JmR -/+) HC cases for weak (BQC = 0.05) and strong (BQC = 0.5) interactions.

The implications behind this discovery could mean that BQC can potentially be leveraged to help stabilize the magnetization of the electrodes long-term for different temperatures if various parameters such as BQC and HC are optimized using techniques such as machone learning.

It appears that under the antiparallel and parallel cases, the leads want to align perpendicular with respect to one another

## 4 Conclusion

When only BQC exists, it can be deduced that BQC does not influence the electrode spins to orient in a preferred orientation as it has no preference for parallelity; the strength at which the electrodes align in a parallel or antiparallel is also on the weaker side, which indicates that BQC weakly influences electrode spin in the electrodes.

Introducing and increasing the BQC strength up to the same magnitude as HC was found to have little influence on device magnetization; it only assisted the device in achieving magnetic equilibrium more quickly for the device with extended cross junctions as shown in the comparison between the evolution between BQC = 0/0, BQC = 0.5/0.5 and BQC = 1/1 shown in Figs. 5(A), 5(B) and 5(C), respectively. The spins of the molecules only try to establish a 90-degree orientation with respect to the left and right electrodes, especially at the highest BQC strength (bmL/bmR = 1/1).

Both the MTJMSD cross junction and pillar simulations exhibit the same FM behavior regardless of increased BQC strength, as shown in the temporal evolution graphs in Fig. 5; however, the pillar MTJMSD reach stability much quicker than the extended cross junctions. This can be attributed to the fact that the extended cross junction has one order of magnitude more molecules to align as compared to the pillar.

Curiously, there exists high correlation between the molecule spins and the spins of the left and right electrodes near the molecules for all BQC strengths examined (bmL/bmR = 0/0 to bmL/bmR = 1/1), but some portions of the extended edges have lower correlation with respect to the rest of the device. This resistance against uniformly orienting the spins at the edges with the spins of the rest of the electrode body could potentially be explained by the idea that the effect of BQC is a predominantly localized behavior and is competing with the magnetic moments of the additional magnetic mass from the extended leads.

The difference between the two configurations' magnetic behavior lies in the degree of stability to which the molecules of the left and right electrodes are oriented antiparallel with respect to one another. There is a greater resistance in the pillar device compared to the extended cross junction which posits that the extended leads may have the effect of helping to stabilize the device when under the influence of high molecular HC.

Results confirm that BQC has a weaker exchange coupling compared to HC, which means that HC ultimately dominates the MTJMSD's magnetic behavior; however, rather than competing with HC, BQC was shown to assist in stabilizing the overall equilibrium magnetic moment of the device. If the device has strong parallel (antiparallel) molecular HC, the device will continue to maintain its ferromagnetic (antiferromagnetic) behavior, regardless of the introduction of strong molecular BQC in the device. Further investigation of the competing effects of HC and BQC can be done through the conduction of a controlled experiment of these devices. Further examination of BQC and HC in a larger domain such as the actual electrodes is also another condition that could be explored.

Temperature

This behavior can be explained by the fact that additional thermal energy will excite the electrons to randomly orient themselves until they reach a spin state that is energetically favorable to maintain (i.e., the antiferromagnetic state). This also means that the increased thermal energy reduces the spins' resistance to altering their spin state and will be able to switch state easily. This has important implications when it comes to understanding what temperature the devices should be subjected to in order to ensure that they exhibit desired magnetic behavior for various purposes. For example, in memory devices such as MRAMs, it is desired to have the device's spins to be maintained for as long as possible. Finding an optimal temperature range for this type of scenario then is necessary.

## Acknowledgements

This paper was prepared under the MECH 500 Research Methods and Technical Communication course taught by Professor Pawan Tyagi in the Fall of 2022. We acknowledge funding support for this course from National Science Foundation-CREST Award (Contract # HRD- 1914751), the Department of Energy/ National Nuclear Security Agency (DE-FOA-0003945), and the NASA MUREP grant (80NSSC19M0196)

## References

1. Kartsev, A., et al., *Biquadratic exchange interactions in two-dimensional magnets.* npj Computational Materials, 2020. **6**(1).
2. Tyagi, P., C. Baker, and C. D'Angelo, *Paramagnetic molecule induced strong antiferromagnetic exchange coupling on a magnetic tunnel junction based molecular spintronics device.* Nanotechnology, 2015. **26**(30): p. 305602.
3. Tyagi, P., et al., *Molecular coupling competing with defects within insulator of the magnetic tunnel junction-based molecular spintronics devices.* Sci Rep, 2021. **11**(1): p. 17128.
4. Tyagi, P., *Molecular Spin Devices: Current Understanding and New Territories.* Nano, 2011. **04**(06): p. 325-338.
5. Maciel, N., et al., *Magnetic Tunnel Junction Applications.* Sensors (Basel), 2019. **20**(1).
6. Demokritov, S., *Biquadratic interlayer coupling in layered magnetic systems.* 1997.
7. Rührig, M., et al., *Domain Observations on Fe□Cr□Fe Layered Structnres. Evidence for a Biquadratic Coupling Effect.* Physica Status Solidi (a), 1991. **125**(2): p. 635-656.
8. Heinrich, B., et al., *Bilinear and biquadratic exchange coupling in bcc Fe/Cu/Fe trilayers: Ferromagnetic-resonance and surface magneto-optical Kerr-effect* *studies.* Phys Rev B Condens Matter, 1993. **47**(9): p. 5077-5089.
9. Slonezewski, J., *Fluctuation Mechanism for Biquadratic Exchange Coupling in Magnetic Multilayers.* Physical Review Letters, 1991.
10. Layadi, A., *Effect of biquadratic coupling and in-plane anisotropy on the resonance modes of a trilayer system.* Physical Review B, 2002. **65**(10).
11. Gareev, R., *Antiferromagnetic interlayer exchange coupling across epitaxial, Ge-containing spacers.* American Institute of Physics, 2003. **83**.
12. Layadi, A., *Study of the resonance modes of a magnetic tunnel junction-like system.* Physical Review B, 2005. **72**(2).
13. Layadi, A., *Analytical expressions for the magnetization curves of a magnetic-tunnel-junction-like system.* Journal of Applied Physics 2006. **100**(8).
14. Brown, H., et al., *Impact of direct exchange coupling via the insulator on the magnetic tunnel junction based molecular spintronics devices with competing molecule induced inter-electrode coupling.* AIP Advances, 2021. **11**(1).
15. Tyagi, P., *Magnetic Tunnel Junction Based Molecular Spintronics Devices Exhibiting Current Suppression.* Organic Electronics, 2019. **64**(1566-1199): p. 188-194.
16. Grizzle, A., C. D’Angelo, and P. Tyagi, *Monte Carlo simulation to study the effect of molecular spin state on the spatio-temporal evolution of equilibrium magnetic properties of magnetic tunnel junction based molecular spintronics devices.* AIP Advances, 2021. **11**(1).
17. Newman, M., *Monte Carlo Simulations in Statistical Physics.* 1999.
18. Katzgraber, H., *Introduction to Monte Carlo Methods.* 2011.
19. Joshi, V.K., *Spintronics: A contemporary review of emerging electronics devices.* Engineering Science and Technology, an International Journal, 2016. **19**(3): p. 1503-1513.